\documentclass[aps,longbibliography,preprint,showkeys]{revtex4-1}

\usepackage{babel,rotating,dcolumn}
\usepackage{colordvi,graphicx,color,amsbsy,amsmath,bm,amssymb}
\usepackage{type1cm}

\begin{document}

\title{On the local anisotropy of quasi-two-dimensional forced shallow flow: an experimental study}

\author{Ghassan Antar$^{1}$, Jamal El Kuweiss$^{1}$, Kai Schneider$^{2}$,  Charbel Habchi$^{3}$ and Sadruddin Benkadda$^{4}$}

\affiliation{$^{1}$ Department of Physics, American University of Beirut, Riad El-Solh, Beirut, 1107 2020, Beirut, Lebanon}

\affiliation{$^{2}$ Institut de Math\'ematiques de Marseille (I2M), Aix-Marseille Universit\'e, CNRS, 3 place Victor Hugo, 13331 Marseille Cedex 3, France}

\affiliation{$^{3}$ Mechanical Engineering Department, Notre Dame University-Loueize, Zouk Mosbeh, Zouk Mikael, P.O.Box 72, Lebanon}

\affiliation{$^{4}$ PIIM, Aix-Marseille Universit\'e, CNRS, Av. Escadrille Normandie Niemen, 13397 Marseille Cedex 20, France}


\begin{abstract}
We experimentally investigate quasi-two-dimensional (Q2D) forced shallow flows in the presence of solid boundaries and analyze the deviation from the Kolmogorov-Kraichnan (KK) theory. Complex motion is generated using a thin electrolyte subject to electromagnetic forces, and we employ particle tracking velocimetry to resolve the flow properties down to the Kolmogorov scale. Although the velocity probability distribution function closely resembles a Gaussian, deviations from Gaussianity emerge for velocity increments as scales decrease. The second-order structure function supports the onset of local anisotropy at small scales. The sign of the third-order structure function indicates the dominance of the inverse cascade in energy transfer, and the cross-correlation between longitudinal and transverse directions proves to be significant at large scales. The breakdown of local isotropy is consistent with the effect of bottom friction, which primarily affects the longitudinal motion, while leaving the perpendicular direction unaffected. This local anisotropy propagates to larger scales {\em via} the inverse energy cascade, with nonlinear interactions eventually influencing the perpendicular direction. 
\end{abstract}


\maketitle

\section{Introduction}

Quasi-two-dimensional (Q2D) turbulence occurs in flows where the dynamics are dominated by motion in two directions. In geophysical fluid dynamics, the Earth's atmosphere and oceans have thicknesses much smaller than the Earth's radius; thus, studying large-scale atmospheric and oceanic circulation patterns often involves two-dimensional approximations. Q2D turbulence helps us understand the dynamics of weather patterns, ocean currents, and climate phenomena such as the formation of jets (the jet stream) and eddies (cyclones, anticyclones, and turbulent vortices). Another way to reduce the role of the third dimension is in magnetically confined plasmas, where perturbations in the direction parallel to the magnetic field are much smaller than in the perpendicular plane~\citep{diamond2010modern}. Rotating or stratified flows such as atmospheric circulation, ocean currents, and galactic rotations also exhibit a reduction in dynamics in one dimension~\citep{alexakis2023quasi,narimousa1991experiments}.

Three-dimensional (3D) turbulence exhibits non-predictable rotational motion occurring in three dimensions. The theoretical work of A. N. Kolmogorov thought of this complexity as caused by an energy cascade. Assuming a large Reynolds number, local isotropy, and self-similarity, he obtained his famous 2/3-law prediction~\citep{kolmogorov1941local,kolmogorov1962refinement}. Kraichnan applied similar arguments to 2D turbulence, where not only the turbulent kinetic energy (the average of the velocity square) is conserved but also the enstrophy (the average of the vorticity square)~\citep{kraichnan1967inertial}. This led to the prediction of the co-existence of two cascades on either side of the forcing scale. In the direct cascade, enstrophy is transferred towards smaller scales, while in the inverse cascade, energy moves up to larger scales.

Two main types of laboratory experiments were elaborated to understand the dynamics of Q2D flows~\citep{sommeria1986experimental, kellay2002two,clercx2009two,  boffetta2012two}. The first type is gravity-driven soap films, where the thickness is about 100~$\mu$m, and turbulence is generated by a grid~\citep{gharib1989liquid,belmonte1999velocity}. Electromagnetically driven flows, on the other hand, use a set of permanent magnets installed underneath the container, and a current is driven between the electrodes~\citep{paret1997experimental,boffetta2005effects,shats2005spectral}. In this case, the thickness may be modified from one to several millimeters, and turbulence is generated by the nonlinear interaction among the vortices. The statistical properties of Q2D flows obtained by the different groups do not converge on whether the laboratory flows obey the Kolmogorov-Kraichnan theory or not. Some have indicated a good agreement~\citep{tabeling1991experimental,xia2008turbulence}, but others did not~\citep{chen2006physical,boffetta2012two}. 

The existence of the condensate, which is one large steady vortex, may be one of the reasons why the experiment disagrees with the theory~\citep{xia2008turbulence}. A strong condensate leaves a footprint on the underlying turbulence as it reduces the randomness of the flow and reduces the efficiency of the inverse energy cascade~\citep{xia2009spectrally}. In a bounded flow at low bottom dissipation, the inverse energy cascade leads to the generation of a spectral condensate below the free surface. Such a coherent flow can destroy 3D eddies in the bulk of the layer and enforce flow planarity throughout the thickness of the layer~\citep{byrne2011robust}. A quantitative study of the turbulent diffusion shows a significant decrease of the radial transport during the spectral condensation process~\citep{bardoczi2012inverse}.

The other major phenomenon is the solid no-slip boundary at the bottom of the container, which could lead to a profound deviation from the KK theory developed for 2D turbulence. It was argued that the usual assumption that a shallow fluid flow is Q2D is wrong and that there are still some 3D effects. These effects were shown to be not due to the bottom drag but to the impermeability of the boundaries~\citep{akkermans2008three}.

This paper is dedicated to the experimental investigation of a Q2D flow that is electro-magnetically driven by two electrodes and a set of permanent magnets. We wish to investigate the effects of the no-slip solid boundary on the statistics of the velocity fluctuations. Thus, no intermediate fluid layer is employed between the solid bottom and the electrolyte. We characterize the flow motion using particle tracking velocimetry (PTV), where 50~$\mu$m fluorescent particles are initially puffed on the surface and followed by visible imaging. We continue with the description of our experimental setup and the diagnostic used in Sec.~\ref{sec-setup}. In Sec.~\ref{sec-random}, we present the effect of increasing the current and achieving a turbulent regime. We emphasize that, throughout this paper, a clear distinction is made between a `turbulent regime' and a `fully developed turbulent regime'. In the former, our case, the flow exhibits complex and irregular dynamics with strong nonlinear interactions, yet some degree of spatial or temporal coherence may persist, as is often observed in transitional or partially turbulent flows. In contrast, a fully developed turbulent regime corresponds to conditions at sufficiently high Reynolds numbers, where turbulence becomes statistically homogeneous and the flow is characterized by a broad and well-separated range of interacting scales, from the largest energy-containing eddies down to the smallest dissipative structures. The probability distribution function of the velocity and its increments are discussed in Sec.~\ref{sec-pdf}, showing a deviation from a Gaussian distribution at small scales. The second-order structure functions in the laboratory and moving frames of reference are presented in Sec.~\ref{sec-S2}, which is followed by Sec.~\ref{sec-S3} discussing the third-order structure function. Conclusions are drawn in Sec.~\ref{sec:concl}, where we discuss the role of the bottom no-slip boundary, which allows friction to break the local isotropy of the flow at small scales. Larger scales are affected by the inverse cascade, which is shown to dominate energy transfer. The strong cross-correlation between longitudinal and transverse motion contributes to the deviation of not only the longitudinal but also the transverse moments from the KK theory.


\begin {figure}[h]
\centerline{\includegraphics[scale=0.5]{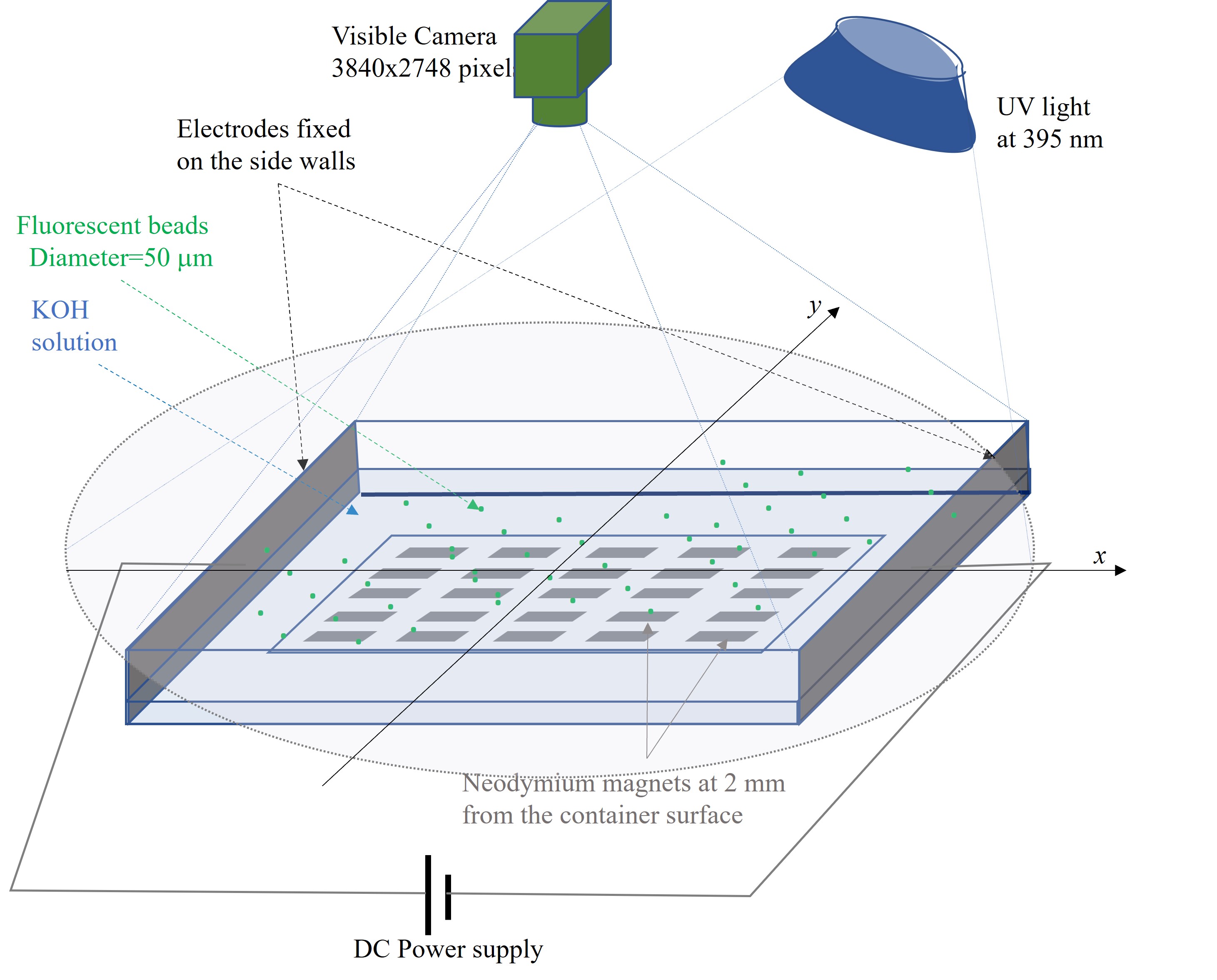}}
\caption{The experimental setup is illustrated, showing the container with two rectangular electrodes connected to a variable DC power supply. Beneath the container, we insert a set of permanent magnets with opposite polarities. The green dots represent the fluorescent beads deposited on the flow surface. Ultraviolet light sources illuminate the flow, and a camera detects the visible light emitted by the beads.}
\label{fig:setup} 
\end{figure}

\section{The Experimental Setup} \label{sec-setup}

\subsection{The fluid setup}

\begin {figure}[h]
\centerline{\includegraphics[scale=0.5]{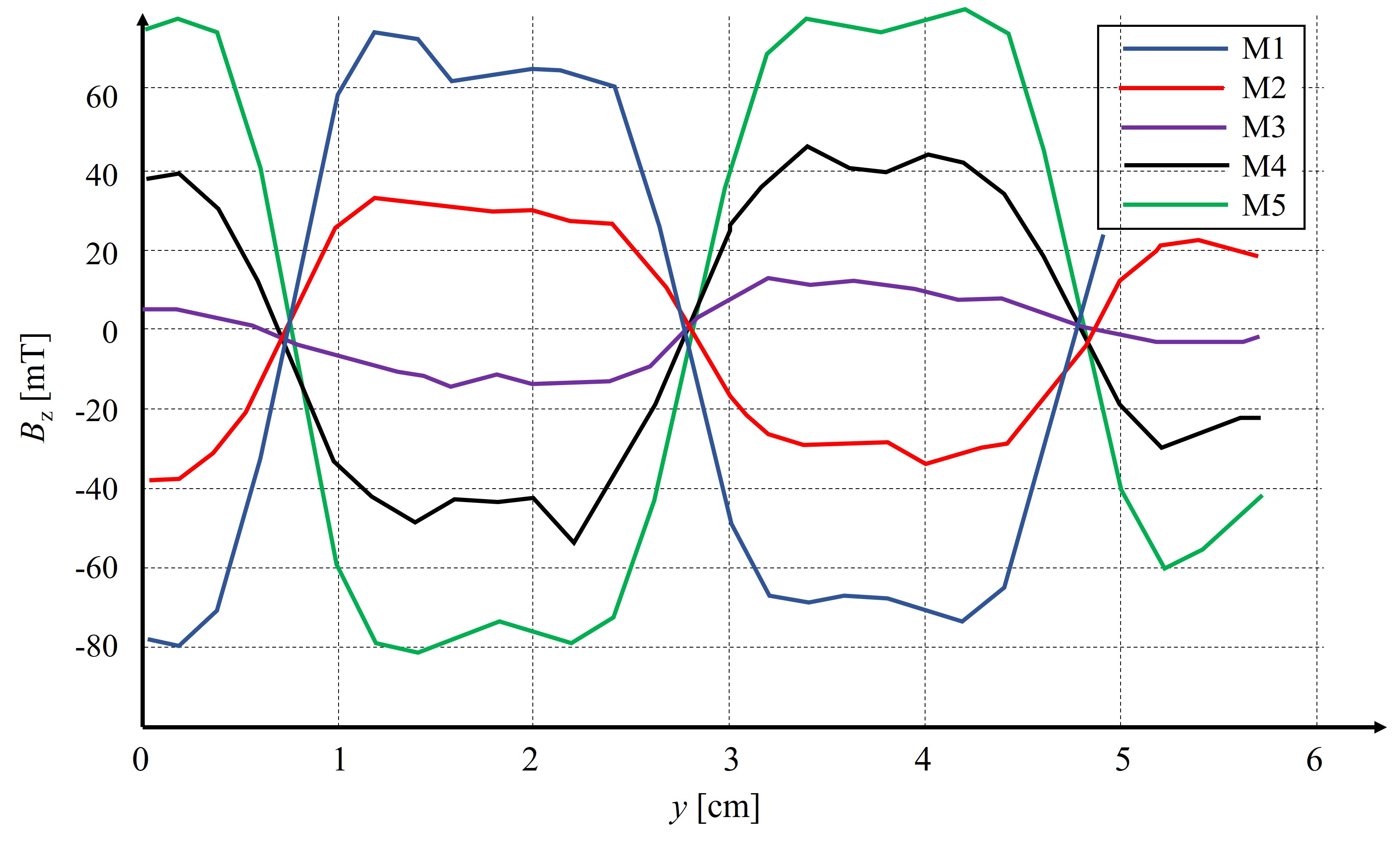}}
\caption{The Teslameter, positioned at 5~mm from the magnet top surface, yields The $z$-component of the magnetic field, $B_z$, as a function of $y$ at five different $x$ locations, with graphs denoted by M1-M5. } 
\label{fig:Bvsy} 
\end{figure}

The experimental setup is illustrated in Fig.~\ref{fig:setup}. We use a square container made of Plexiglas with an inner length of $W=16$~cm and a height equal to 2~cm. Two stainless steel electrodes are mounted on two opposing walls and connected to a variable DC power supply. Pure water is mixed with potassium hydroxide (KOH) to reach a concentration of 26$\%$. This generates positively (K$^+$) and negatively (OH$^-$) charged ions that are attracted and repelled, respectively, at the cathode and anode surfaces. The conductivity of this electrolyte solution is 55~S/m, which is about $10^{-6}$ that of copper. Consequently, this solution remains an insulator in terms of electron mobility, and the resultant measured current is due to the charge exchange at the electrode surface without a net flow of electrons inside the solution. The amount of KOH solution poured into the container defines the height $H$ of the flow and can be modified. The setup is adequately leveled so that all parts of the fluid are subject to the same gravitational force. A set of $6\times5$ permanent neodymium magnets is installed under the container with opposite polarities, leading to a magnetic field gradient of about 28~T/m as can be deduced from Fig.~\ref{fig:Bvsy}, where the axial magnetic field is plotted as a function of $y$.

The kinematic viscosity of water is used to assess the Reynolds number, defined as $Re = UD/\nu$. Because there is no mean flow, we use the root-mean-square value of the velocity, leading to $U \simeq 2$~cm/s. The typical macroscopic scale of turbulence in the horizontal plane is taken to be the distance between magnets $L\simeq 2$~cm. In Table~\ref{tab:my_label}, we insert some important parameters of our flow with $I$ being the current drawn between the electrodes.

\begin{table}[h]
   \centering
    \begin{tabular}{|c|c|c|c|c|}
    \hline
        {\it I} [A] & $H$ [mm] & $H/W$ & $H/L$ & $Re=UL/\nu$   \\ \hline
        0.1 & 3 & 0.019 & 0.15 & 100 \\ \hline
        0.5 & 3 & 0.019 & 0.15 & 200 \\ \hline
        0.7 & 5 & 0.03 & 0.25 & 180  \\
     \hline   
    \end{tabular}
    \caption{The main flow parameters for the two heights and two currents investigated here. $L$ is the distance between the magnets.}
    \label{tab:my_label}
\end{table}

\subsection{Particle tracking velocimetry}

To obtain the velocity field as a function of space and time, we scatter fluorescent beads (from Cospheric), with density 0.98~g/cm$^3$ and diameter 38-45~$\mu$m on the surface of the fluid. They absorb ultraviolet (UV) light and emit it in the visible range. To excite them, we install two LED spotlights that emit around 395~nm with a total power of 30~W, at approximately 40~cm from the electrolyte solution. The visible camera is a Basler ACA3800-14uc that has a full resolution of $3840\times 2748$~pixels at 14 frames per second. It is equipped with a 50~mm C-mount lens that allows efficient coverage of the setup and thus a spatial resolution of $42\times58$~$\mu$m. Consequently, each bead occupies about one pixel in the camera image, and this considerably reduces the error in determining their position and thus their velocity.

After pouring the KOH solution into the container, we scatter approximately 500 beads onto its surface. This quantity represents a balance: too many beads would impede the accurate determination of their velocity field via particle tracking velocimetry (PTV), while too few would decrease the statistical reliability of the results. The motion of the beads is recorded at 1,000 frames (corresponding to a 1-minute movie), generating roughly 500,000 data points per movie. To enhance the statistics, ten movies are recorded for each experimental condition. The total number of data points used in this paper is 3,219,310 for a height $H = 3$~ mm and 5,812,925 for $H = 5$~mm.

\begin {figure}[h]
\centerline{\includegraphics[scale=0.3]{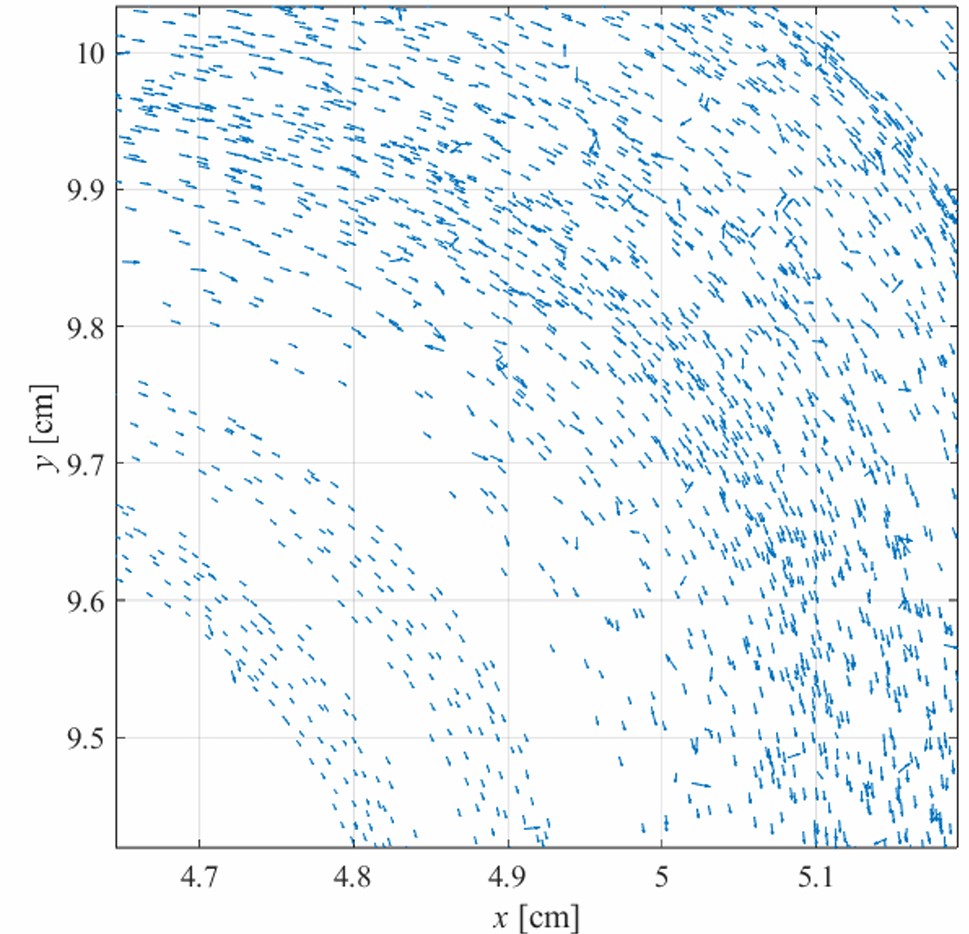}}
\caption{The velocity vector in a region of about $1\times1$~cm$^2$ using arrows for $H=3$~mm and a current of 100~mA. }
\label{fig:quiver} 
\end{figure}

Figure~\ref{fig:quiver} shows a zoomed-in view of a 1~cm$^2$ region. The bead velocities, determined from 100 frames, are overlaid in this image. For a current of 100~mA, the figure reveals the coherent rotational motion, showing a clear clockwise rotation.

\subsection{The determination of the velocity field and its increments}

Each image from the visible camera is processed by first subtracting the background. Beads positions are identified as local intensity maxima. To exclude contributions from agglomerated beads, we filter out all sources with a spatial extent larger than $2\times2$~pixels. In the laboratory reference frame, each bead $i$ in frame $j$ has a position $(x_{i,j},y_{i,j})$ in the $\hat{x}$ and $\hat{y}$ directions. The velocity of each bead is computed using a central difference scheme, which tracks its motion across three consecutive frames. This establishes a smallest resolved spatial scale of $\lambda_{\text{min}} \approx 100$~$\mu$m and a smallest velocity of $u_{min} \simeq 0.07 $~cm/s.
$$  
u_{i,j} = \frac{x_{i,j+1} - x_{i,j-1} }{2\delta t},
\\
v_{i,j} = \frac{y_{i,j+1} - y_{i,j-1} }{2\delta t}.
$$ 
The maximum tracking distance is constrained to the average inter-bead distance to minimize erroneous connections to false neighbors. This constraint sets the maximum resolvable velocity at $u_{\text{max}} \simeq 7$~cm/s. The velocity increments in the laboratory frame are thus defined as:
$$
\delta u(\vec{r}) = u_{i,j} - u_{i',j}, \; \;
\delta v(\vec{r}) =  v_{i,j} - v_{i',j} \, .
$$
where $\vec{r}$ is the vector distance between two beads $i$ and $i'$ in the same frame $j$. 

Knowing both the positions and the velocity fields of the beads allows us to calculate the longitudinal $\delta u_L$ and the transverse $\delta u_T$ velocity increments in the beads' moving frame according to 
$$
\delta u_L = (\vec{u}_{i,j} - \vec{u}_{i',j}) \cdot\hat{r}, \;
\delta u_T =  (\vec{u}_{i,j} - \vec{u}_{i',j}) \cdot\hat{t} \, .
$$
where $\cdot$ is the scalar product, $\hat{r}=\vec{r}/r$ and $\hat{t} \cdot \hat{r}=0$.

\subsection{The Characteristic Scales}

Scales in two-dimensional turbulence are important since they dictate the statistical properties of the velocity fluctuations. The Kolmogorov scale is the smallest length scale associated with the smallest eddies in a turbulent flow. It is defined as $\lambda_K =( \nu^3/\varepsilon)^{1/4}$  where $\varepsilon$ is the turbulent dissipation rate per unit mass. In our flow, it is $\lambda_K \sim 63$~$\mu$m, which is about the spatial resolution of our particle tracking method. We define the `Q2D range' to represent the scales that are greater than $\lambda_K$ and smaller than $H$, thus ranging from $0.06$ to $3-5$~mm. In this limited range, turbulent 3D motion may be present, thus affecting the properties of the flow.

\begin {figure}[h]
\centerline{\includegraphics[scale=0.5]{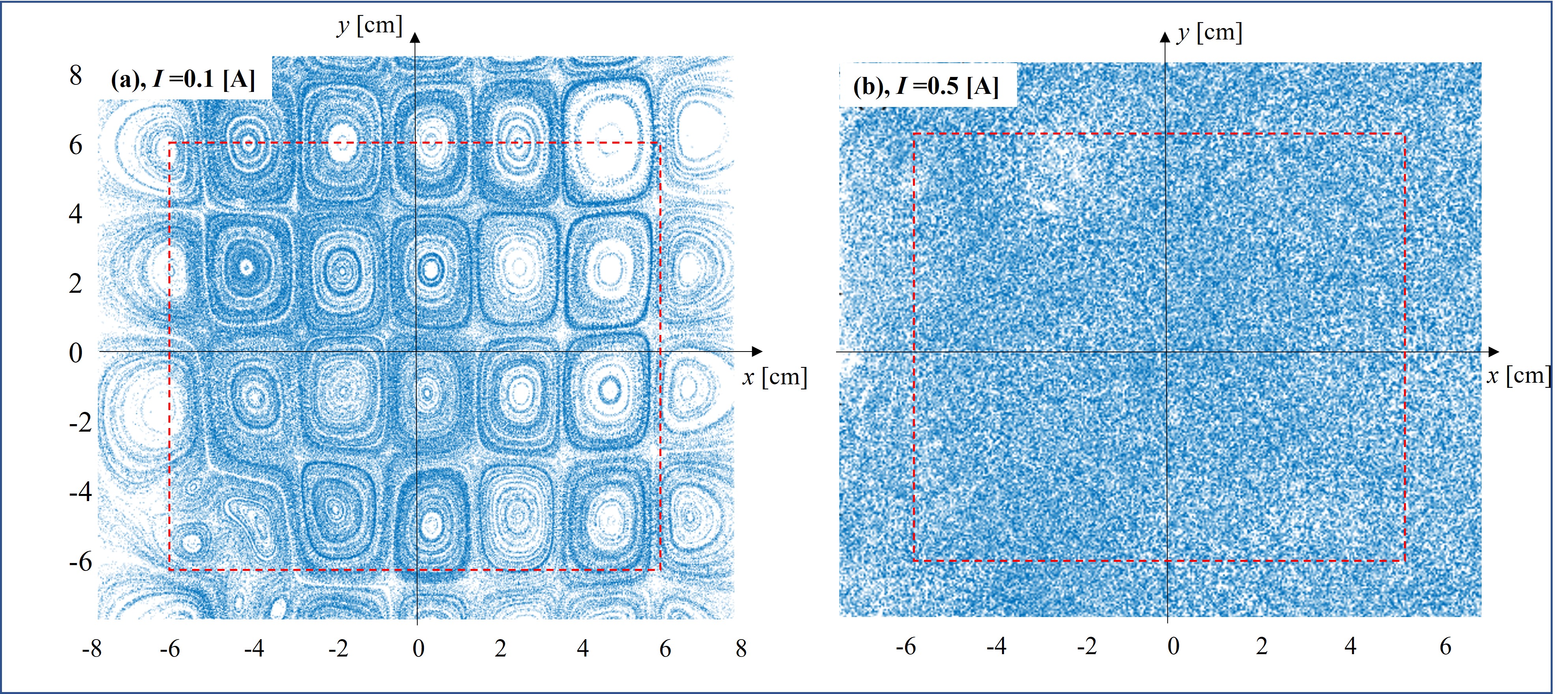}}
\caption{Panels (a) and (b) display the positions of fluorescent beads for a height $H=3$~mm at currents 
$I=100$ and $I=500$~mA respectively. The beads' positions are overlaid across 100 consecutive frames. The dashed rectangle marks the spatial domain used for statistical analysis.}
\label{fig:spatial_randomness} 
\end{figure}

Vortices are generated at a characteristic scale of $L\simeq2$~cm, as evident in Fig.~\ref{fig:spatial_randomness}(a) under low current conditions. According to the Kolmogorov-Kraichnan (KK) theory~\citep{kraichnan1967inertial,lesieur2008introduction}, the scales between $H$ and $L$ should fall within the direct cascade range where energy flows from $L$ toward smaller scales down to $\lambda_K$. In our experiment, we have one decade in this range from 0.3 to 3~cm. For scales greater than $L$, their maximum is the size of the container ($W=16$~cm). We thus have half a decade in this range identified to be the inverse cascade range according to the KK theory.

\section{Spatial structures at low and high currents} \label{sec-random}

Before analyzing the statistical properties of turbulence, we examine the spatial properties of the flow. Fig.~\ref{fig:spatial_randomness} displays the superposition of the bead positions in 100 consecutive frames for two different currents: (a) 100~mA, and (b) 500~mA, at a fixed height $H = 3$~mm.

At $I=100$~mA (Fig.~\ref{fig:spatial_randomness}(a)), we observe well-defined coherent vortices formed by the forcing from magnetic field gradients around each magnet, which produces rotational acceleration~\citep{moubarak2012dynamics}. The forcing is isotropic in the ($x,y$)-plane, showing no directional preference. Here, isotropy implies that the statistical properties remain unchanged under rotation of the reference frame. This differs from the local isotropy definition in the Kolmogorov theory~\citep{kolmogorov1941local}, which requires invariance under both translation and rotation in the {\em moving} reference frame.

At the current equals 500~mA (Fig.~\ref{fig:spatial_randomness}(b)), the bead distribution reveals a random fluid motion, with the complete disappearance of the coherent structures. Moreover, we do not observe vortex condensation into one or several large-scale structures, consistent with the findings of \citet{paret1998intermittency}. With increasing current, we thus record a transition from a flow dominated by coherent vortices to a flow without, which is described as turbulent.

\section{Probability distribution functions of velocity and its increments} \label{sec-pdf}

The probability distribution functions (PDFs) of the velocity field and its increments allow us to quantify the randomness of the motion presented above. Gaussian random fields have skewness and flatness factors that are equal to 0 and 3, respectively.  The deviation of the PDFs from a Gaussian distribution is caused by non-random events that reflect a deviation from the Kolmogorov theory. 

\begin{figure}[h]
\centerline{\includegraphics[scale=0.4]{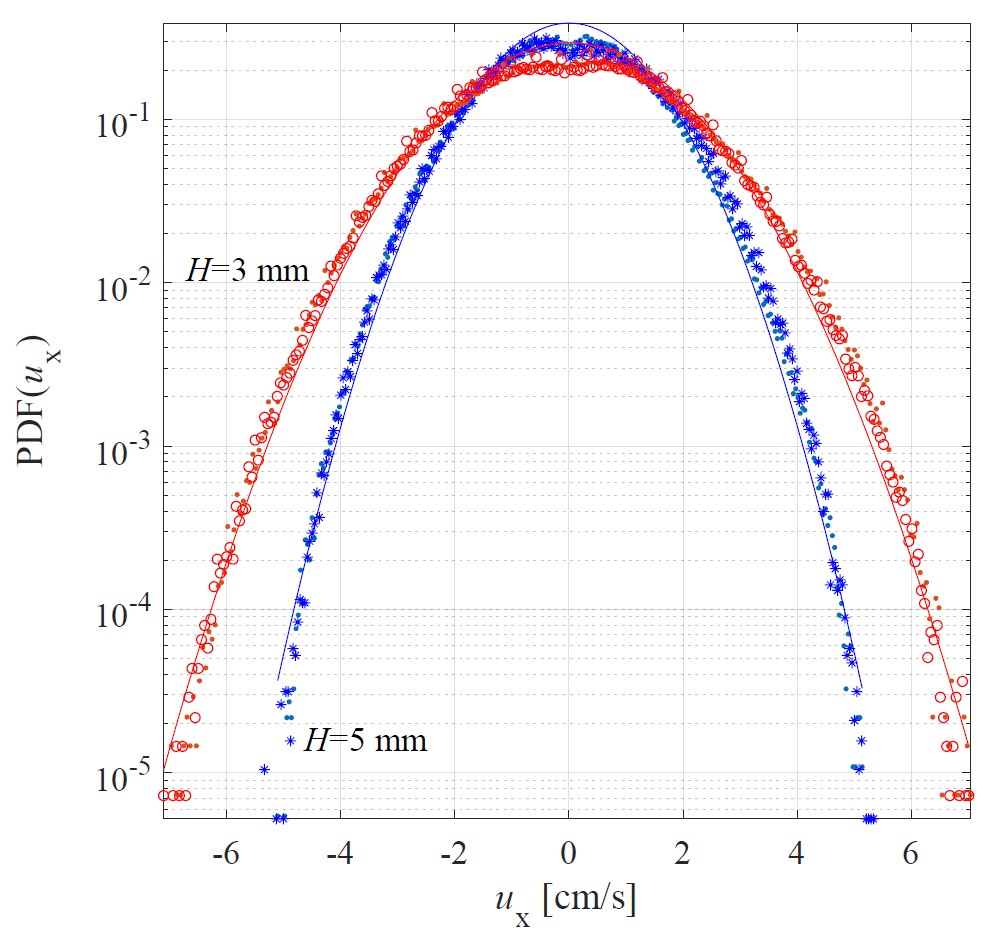}}
\caption{The PDF of the $x$-component of the velocity for $H = 3~\mathrm{mm}$ and $H = 5~\mathrm{mm}$ are shown with circles and stars markers respectively. The solid lines represent the corresponding Gaussian distributions obtained through least-squares fitting. The measured distributions exhibit near-zero skewness ($\approx 0$) with flatness factors of 2.7 ($H = 3~\mathrm{mm}$) and 2.8 ($H = 5~\mathrm{mm}$), indicating slight deviations from Gaussian statistics.}
\label{fig:PDF35} 
\end{figure}

\subsection{PDFs of velocity}

The PDF of the $x$-component of the velocity field is plotted in Fig.~\ref{fig:PDF35} for $H=3$ and 5~mm. The two distributions are fitted with Gaussian functions by applying the least-squares minimization. The PDF slightly deviates from a Gaussian distribution at small scales, most probably caused by using a centered difference to determine the velocity field. To quantify the deviation of the PDFs from Gaussian statistics, we compute the normalized third and fourth-order moments of the velocity fluctuations, known as the skewness and flatness factors. The skewness, defined as $S_x = \langle u^3 \rangle/\langle u^2 \rangle^{3/2}$, yields $S_x = 0$ for both heights, confirming the symmetry of the distribution about $u = 0$. The flatness factor $F_x = \langle u^4 \rangle/\langle u^2 \rangle^2$ measures 2.7 ($H = 3~\mathrm{mm}$) and 2.8 ($H = 5~\mathrm{mm}$), slightly below the Gaussian value of 3. This discrepancy arises from a deficit of high-intensity velocity fluctuations compared to Gaussian predictions, as visible in Fig.~\ref{fig:PDF35} where the tail probabilities fall below the theoretical distributions. However, these deviations remain small and affect only rare events (occurring with probabilities three orders of magnitude below the bulk). We therefore conclude that the velocity field follows approximately Gaussian statistics, consistent with the random motion observed in Fig.~\ref{fig:spatial_randomness}.

\subsection{The PDFs of the velocity increments}

\citet{paret1998intermittency} generated turbulence in electromagnetically driven flows, using thin, but stably-stratified layers to minimize the bottom friction. The PDFs of the longitudinal increments were overlaid for separations in the inverse cascade range, and they were found to remain Gaussian. The authors deduced that this indicates a lack of intermittent events in the inverse cascade range. On the other hand, PDFs were also discussed by \citet{belmonte1999velocity} for two-dimensional flows in soap films. They found that the PDF of $\delta u_L(r)$ becomes non-Gaussian with decreasing $r$, similar to the 3D-turbulence case, indicating a deviation from the random perturbation assumption of Kolmogorov at small scales. 

\begin {figure}[h]
\centerline{\includegraphics[scale=0.6]{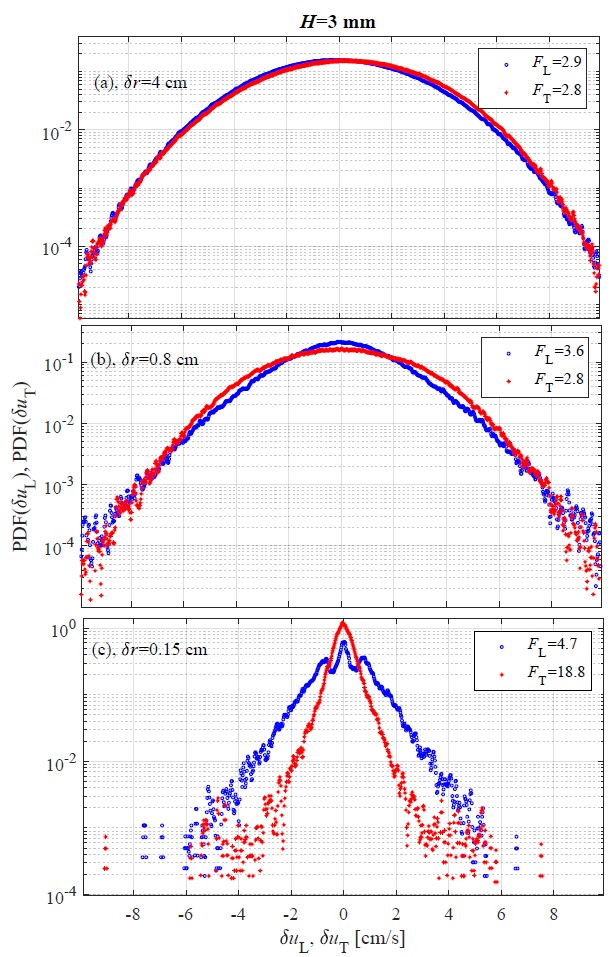}}
\caption{For $H=3$, we plot the PDF of $\delta u_L$ and $\delta u_T$ using three separation distances $r=0.1$, 0.8, and 4~cm in (a), (b) and (c). The corresponding flatness factors are given in each subplot.}
\label{fig:pdfdelta} 
\end{figure}

In this section, we investigate the PDFs of the velocity increments $\delta u_L(r)$ and $\delta u_T(r)$. The basic idea behind this analysis is to test the randomness of the fluctuations on the scale defined by $r$. If the motion of particles is random, then this should also yield a Gaussian random distribution for the velocity increments.  

In Fig.~\ref{fig:pdfdelta}, we plot the PDFs of the velocity increments for three values of $r$ that lie in the Q2D range ($r=0.15$~cm), in the direct cascade range ($r=0.8$~cm), and in the inverse cascade range ($r=4$~cm). The results are shown for $H=3$~mm, but similar results are obtained with $H=5$~mm.

Fig.~\ref{fig:pdfdelta}(a) displays the PDFs of both longitudinal ($\delta u_L$) and transverse ($\delta u_T$) velocity increments for separation distances $r=4$~cm within the inverse cascade range. The two distributions are almost identical, reflecting local isotropy or isotropy in the beads' frame of reference. Both PDFs closely follow Gaussian distributions, as confirmed by their flatness factors, $F_L=\langle \delta u_L^4 \rangle/\langle \delta u_L^2\rangle^2$ for the longitudinal direction ({\it idem} in the transverse direction). The recorded values, $F_L=2.9$ and $F_T=2.8$, are close to 3, the Gaussian value. The skewness factors of the velocity increments in the two directions are equal to 0.

In the direct cascade range, where $r$ equals 0.8~cm, the results are shown in (b). We observe a net deviation from a parabolic shape towards an exponential one for the longitudinal direction, while the transverse one remains close to parabolic. The local isotropy reported in the inverse cascade is thus broken in the direct cascade range. The flatness factors are equal to 2.8 for $\delta u_T$ and 3.6 for $\delta u_L$, which indicates a net deviation from a Gaussian distribution in the longitudinal direction but not in the transverse one. The difference in the behavior in the two directions reflects the lack of local isotropy at this scale. 

For $r=0.15$~cm, which lies in the Q2D range, the anisotropy between the two directions is even more pronounced as shown in Fig.~\ref{fig:pdfdelta}(c). At these scales, both the longitudinal and transverse directions are far from Gaussian with clear exponential tails. This is quantified by the flatness factor reaching $F_L=4.7$ and $F_T=18.8$, indicating an even bigger difference in the two directions. The origin of the two peaks at low velocities in the longitudinal direction is unknown.

We conclude this section by highlighting the facts: (1), the PDFs of the velocity fields are close to a Gaussian distribution, reflecting the randomness visually inspected in Sec.~\ref{sec-random}. (2), at large scales, the PDFs of the velocity increments remain close to a Gaussian shape, but with decreasing scales, they present exponential tails. (3), the local isotropy is broken at small increments with the PDFs for $\delta u_L$ being different from that of $\delta u_T$.

\section{Second-order structure functions} \label{sec-S2}

The pioneering theory of \citet{kolmogorov1941local} provided the first predictions for fully developed three-dimensional turbulence through structure functions, building upon the foundational work of 
\citet{de1938statistical}; One can see also \citet{djenidi2022karman} for a more recent work. 
By invoking self-similarity and local isotropy in the inertial range, Kolmogorov derived the characteristic $r^{2/3}$ energy scaling. In contrast, fully developed two-dimensional turbulence conserves both energy and enstrophy, yielding distinct scaling laws: $r^2$ for scales larger than the forcing scale (inverse cascade) and $r^{2/3}$ for smaller scales (direct cascade)~\citep{kraichnan1967inertial}. Crucially, the local isotropy hypothesis implies identical $r$-dependence for both transverse and longitudinal components in \textit{both} cascade ranges of 2D turbulence.

We define the second-order structure functions in the laboratory frame as
\begin{equation}
S_{XX} = \langle \delta u(r)^2 \rangle, \; S_{YY} = \langle \delta v(r)^2 \rangle \, ,
\end{equation}
where $X$ and $Y$ denote the corresponding coordinate directions. The brackets $\langle \cdot \rangle$ denote an average over space and time for the 10 movies. We define the total energy structure function as $S_2=S_{XX}+S_{YY}$. 

In the bead's frame of reference, we define the second-order structure functions as 
\begin{equation}
S_{LL} = \langle \delta u_{L}(r)^2 \rangle, \; S_{TT} = \langle \delta u_{T}(r)^2 \rangle,    
\end{equation}

where $T$ and $L$ denote, respectively, the transverse and longitudinal directions of the bead's motion. 

\subsection{Previous results}

The correlation properties of turbulence as a function of $r$ can be determined using the structure function, as will be done here, or the energy spectra $E(k)$ obtained from the correlation of the velocity field. 
\begin{eqnarray}
S_{LL}(r) & = & \langle [u_L(\mathbf{x} + \mathbf{r}) - u_L(\mathbf{x})]^2 \rangle = 2\langle u_L^2 \rangle-2\langle u_L(\mathbf{x} + \mathbf{r}) u_L(\mathbf{x}) \rangle \\ &= & 
4 \int_0^\infty E(k) \left(1 - \frac{\sin(kr)}{kr}\right) dk
\end{eqnarray}

If $S_{LL} \sim r^\alpha$, then $E(k) \sim k^{-(\alpha+1)}$. For measurements taken as a function of time, the link to the spatial dependence is done using Taylor's frozen turbulence hypothesis, where the angular frequency is $\omega=Uk$, leading to $E(\omega) \sim E(k)$~\citep{monin2013statistical,frisch1995turbulence}.

\citet{gage1986theoretical} used data taken from airplanes to obtain the energy spectra of atmospheric turbulence and found the $k^{-3}$ and $k^{-5/3}$ scaling for respectively the large and small scales, reflecting the enstrophy and the energy cascades. Using a soap film driven by gravity,~\citet{gharib1989liquid} showed that the energy spectra, determined in the frequency domain, follow $f^{-\alpha}$, where $\alpha$ can take several values depending on the position of the measurement with respect to the downstream distance to the grid where turbulence is generated. 

\citet{kellay1995experiments,kellay1998vorticity} use the same setup and found two ranges in the frequency domain with two scaling exponents: In the high-frequency range, the slope is about $-3.6$, whereas at low-frequency it is approximately $-2$. However, the authors indicate that because only one-third of a decade was investigated, these measurements cannot be regarded as conclusive. This scaling of the energy was later confirmed by Belmonte {\em et al.} where the slope was found to be equal to $-3.3$~\citep{belmonte1999velocity}. The difference with the value $-5/3$ was argued to be caused by large, long-lived coherent structures and finite-size effects. 

\citet{tabeling1991experimental} performed experiments in thin, stably stratified layers with an electromagnetic force, and showed that the energy spectrum displays a $k^{-5/3}$ dependence, which is consistent with the prediction in the inverse cascade range~\citep{paret1997experimental}. The velocity was measured using particle imaging velocimetry on a grid of $64 \times 64$ in a $15 \times 15 $~cm container~\citep{cardoso1994quantitative}.~\citet{boffetta2005effects}, the energy spectra displayed a power law that is steeper than the Kraichnan prediction $k^{-3}$, which was interpreted to be caused by the bottom friction. 

In 2006~\citet{chen2006physical} published a paper that contains experimental results using an electromagnetically forced layer of salt water 3~mm thick and $18 \times 18$~cm in lateral extents. A heavier 3~mm-thick buffer layer of Fluorine was used to minimize the bottom friction of the forced layer. The velocity fields were obtained using particle tracking velocimetry, resolving velocities on a $100 \times 100$ spatial grid. The energy spectrum did not show an agreement with the theory, the authors suggested that this may be caused by the limited range of scales that was investigated. On the other hand, using a similar setup,~\citet{xia2008turbulence,xia2009spectrally} obtained power laws that are consistent with the KK prediction. They emphasized that the presence of a ``condensate strongly modifies both turbulence level and its statistics; different velocity moments are affected at different scales''. In~\citep{byrne2011robust,xia2011upscale,xia2017two}, the authors show that the power spectra obtained depend on where they are determined; the closer to the bottom, the farther they are from the theoretical prediction, highlighting the effects of the bottom layer. 

\citet{von2011double} measured the horizontal surface flow induced by Faraday waves in a thin fluid layer (2~mm). They obtained an inverse energy cascade with negative mean spectral energy flux and a Kolmogorov-type scaling range. Moreover, the data suggested the existence of a direct enstrophy cascade with a positive mean spectral enstrophy flux. Later, the team presented experimental results that support the existence of the inverse energy cascade fueled by Faraday waves~\citep{francois2013inverse,francois2014three,xia2017two}. 
On the other hand, \citet{bardoczi2012inverse} performed experiments using a NaCl solution with different heights ranging from $2-8$~mm. They reported the existence of a condensate with scales about that of the container size that dominates the turbulence dynamics in steady state. 


\subsection{The second-order structure function in the laboratory frame}

In this section, we present results about the correlation properties in our turbulent flow using the second-order structure functions in the laboratory frame.

\begin {figure}[h]
\centerline{\includegraphics[scale=0.5]{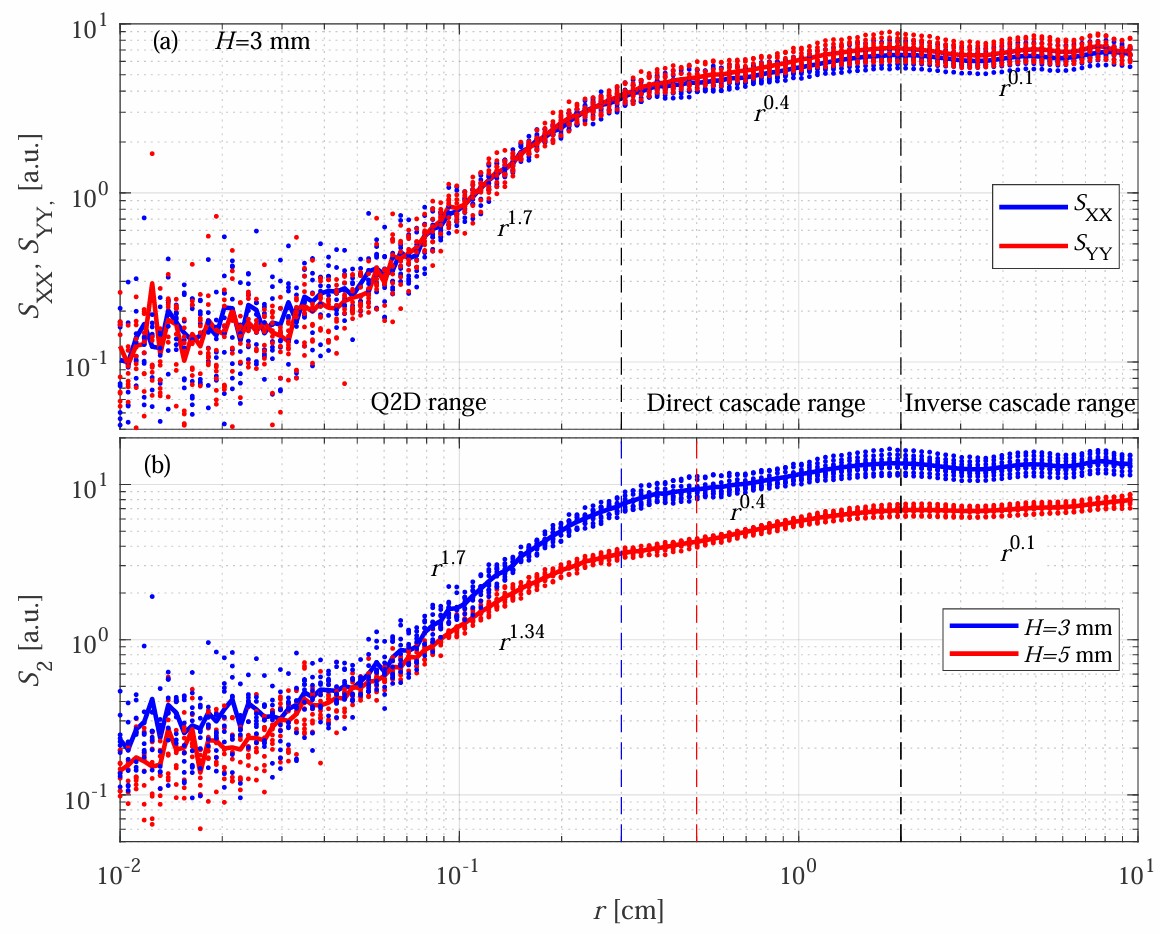}}
\caption{(a), the second-order structure function in the laboratory frame in the $x$ ($S_{XX}$) and $y$ directions ($S_{YY}$) as a function of $r$ for $H=3$~mm. The vertical dashed lines indicate the
position of $r=L$ and $r=H$. The scaling exponent for the best fit by a power law leads to an exponent equal to 1.7 and 1.34, for $H=3$ and 5~mm, respectively. In the direct range, we report the power law $r^{0.4}$, whereas in the direct cascade range, almost no variation is detected with a best fit according to $r^{0.1}$. Note that there is no difference between the two directions reflecting =that turbulence is isotropic. In (b), The total energy spectra, $S_2=S_{XX}+S_{YY}$, is plotted as a function of $r$ for $H=3$, and $H=5$~mm.}
\label{fig:2Euler} 
\end{figure}

In Fig.~\ref{fig:2Euler}(a), we plot $S_{XX}$ and $S_{YY}$ as a function of $r$. The same behavior is recorded, reflecting the isotropy of the flow. This confirms that the statistics of the measured velocity field are sound, and no dependence on the coordinate direction is found as expected.

For $r<H$, the dynamics are in the Q2D range scale like $r^{1.7}$ for the two directions. Between 0.1 and 0.5~cm, we record a saturation that could be caused by viscous dissipation as we approach the Kolmogorov scale.

For $L>r>H$, according to the KK theory, fluctuations should result from a direct cascade with a dependence according to the power law $r^2$. In this range, however, we have a net increase with $r$ with a best fit $r^{0.4}$ (in the wavenumber space $k^{-1.4}$), which is far from the $r^2$ expected from the theory. 

For $W>r>L$, velocity perturbations are caused by an inverse cascade of energy, and the power law expected is $r^{2/3}$. We recall that in the statistically steady-state regime, no condensate is observed. Our experimental results show that in the inverse cascade range, the best fit $r^{0.1}$ ($k^{-1.1}$) is recorded, which is marginally greater than 0, indicating almost no dependence on $r$. 

Fig.~\ref{fig:2Euler}(b) displays the total second-order structure function $S_2$ measured at both fluid heights. In the inverse and direct cascade ranges, the same scaling is obtained within the experimental error. This is not the case in the Q2D range, where the scaling changes from $r^{1.7}$ for $H=3$~mm to $r^{1.34}$ for $H=5$~mm. Note that the current for $H=5$~mm is 700~mA, yet the level of velocity fluctuations remains smaller than for $H=3$~mm. In practice, for $H=5$~mm, a higher current is needed to reach a turbulent state reflected in the absence of coherent structures. 

The behavior of the structure functions in the laboratory frame of reference shows that (1), turbulence is isotropic, (2) the scalings with respect to $r$ deviate from the KK theory at all scales, and (3), the scaling laws depend on $H$, which indicates a lack of universality.

\subsection{The second-order structure functions in the moving reference frame}

We recall that the longitudinal and transverse second-order structure functions are denoted by $S_{LL}$ and $S_{TT}$, respectively, in the directions parallel and perpendicular to $\vec{r}$, which is the vector distance between two beads that belong to the same frame. 
The two directions are related by 
\begin{equation}
S_{TT} = \frac{d}{dr}(rS_{LL}).    
\end{equation}

Consequently, when $S_{LL} \sim r^\alpha$, the same scaling will be obtained for $S_{TT}$.

\begin {figure}[h]
\centerline{\includegraphics[scale=0.4]{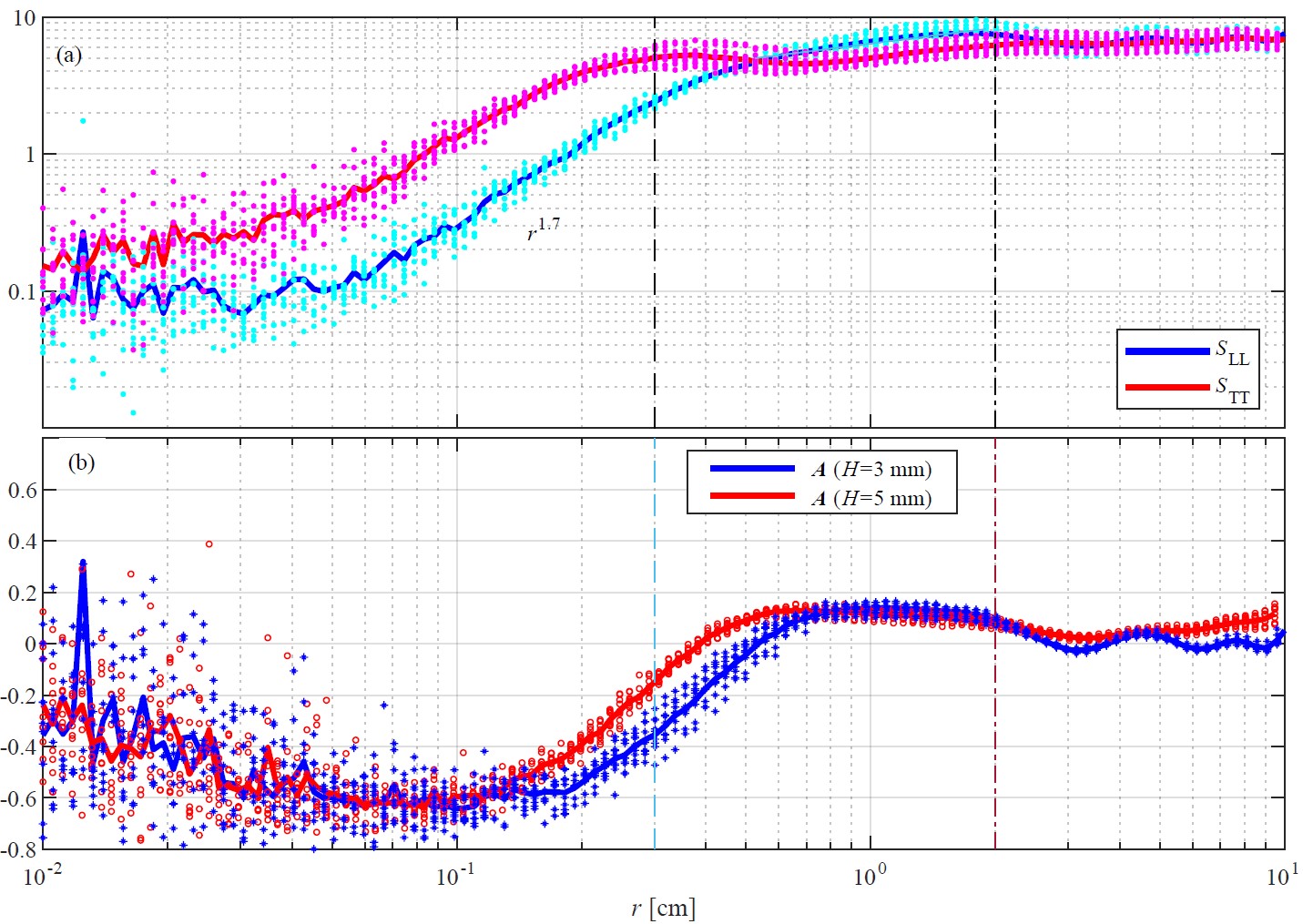}}
\caption{(a), the second-order structure function in the longitudinal ($S_{LL}$) and transverse ($S_{TT}$) directions for $H=3$~mm. In (b), we plot the anisotropy factor ${\cal A}$ as a function of $r$ for the two heights. }
\label{fig:2Lagrange} 
\end{figure}

For $H=3$~mm, the structure functions in the two directions are plotted in Fig.~\ref{fig:2Lagrange}(a) for the 10 movies. This gives an idea about the error bars as a function of $r$.

For the range of scales corresponding to the inverse cascade range ($W>r>L$), the two directions agree, and the best fit indicates $r^{0.1}$. The oscillations detected in the $x$ and $y$ directions appear to be caused by motion in the transverse direction, while they are absent in the longitudinal direction.

In the range $L>r>H$, the best fit indicates a power law $r^{0.4}$, which is in agreement with the power law observed in this range in the laboratory frame. One major difference between the two reference frames is the increase of the structure function around $r\approx H$ in the longitudinal direction, forming a `bump' in the energy spectrum around $H$. We thus record a non-monotonous decrease of the second-order structure function with respect to $r$. This suggests the presence of another source of turbulence at the small scales that generates fluctuations that lead to an accumulation of energy around $H$. 

The power laws in the Q2D range, $\lambda_K<r<H$, in the longitudinal and transverse directions present a scaling exponent of about 1.7, similar to the laboratory frame. The scaling extends in the longitudinal direction over a wider range than in the transverse direction, where $S_{TT}$ remains constant for $r<0.4$~cm. 

The difference in the structure functions' dependence on $r$ in the two directions at small scales is clear, mainly in the Q2D range, where $S_{LL} < S_{TT}$ for $r<H$. This difference appears even at the smallest scales detected by our experiment, which are close to the Kolmogorov scale. This is better represented in Fig.~\ref{fig:2Lagrange}(b), where we plot for the two heights, the anisotropy factor defined as 
\begin{equation}
     {\cal A} = \frac{S_{LL}-S_{TT}}{{S_{LL}}+S_{TT}}.
\end{equation}    
The behavior of ${\cal A}$ is non-monotonous and depends on $r$, which indicates the breaking of the local isotropy hypothesis. Its value increases from $-0.3$, at 0.01~cm, to $-0.6$ at 0.1~cm. Then, it decreases toward 0 and overshoots to a maximum of $+0.17$ for scales around 1~cm where fluctuations in the transverse direction exceed those in the longitudinal one. This plot shows that anisotropy, reflected in the $r$-dependence of ${\cal A}$, is strongest at the small scales, reaching a maximum at $r \approx H/2$. 

We interpret the difference in the small-scale dynamics between longitudinal and transverse as caused by the bottom no-slip boundary, which allows friction to act in the parallel direction to the motion while being absent in the transverse direction. The longitudinal velocity gradient in the $z$-direction would thus generate 3D motion that could explain why $S_{LL} < S_{TT}$ for $r<H$. The presence of another source of turbulence in the Q2D range can also explain the non-monotonous decrease of the longitudinal structure function with $r$ around $r=H$, while this effect is absent in the transverse direction. The results of the second-order structure function show a clear deviation from local isotropy, one of the major assumptions in the KK theory, in agreement with the results of the previous section about the probability distribution function.

\section{Third-order structure functions} \label{sec-S3}

The third-order structure function is associated with the rate of transfer of energy and enstrophy among the different scales through non-linear interactions of velocity fluctuations~\citep{danilov2000quasi,frisch1995turbulence,cerbus2017third,xie2018exact}. The energy transfer rate can be broken down into four parts: $S_{LLL}$, $S_{TTT}$, $S_{LLT}$, and $S_{LTT}$. This section presents the experimental results of the longitudinal and transverse third-order structure functions, and we leave $S_{LLT}$ and $S_{LTT}$ to the next section.

\begin {figure}[h]
\centerline{\includegraphics[scale=0.3]{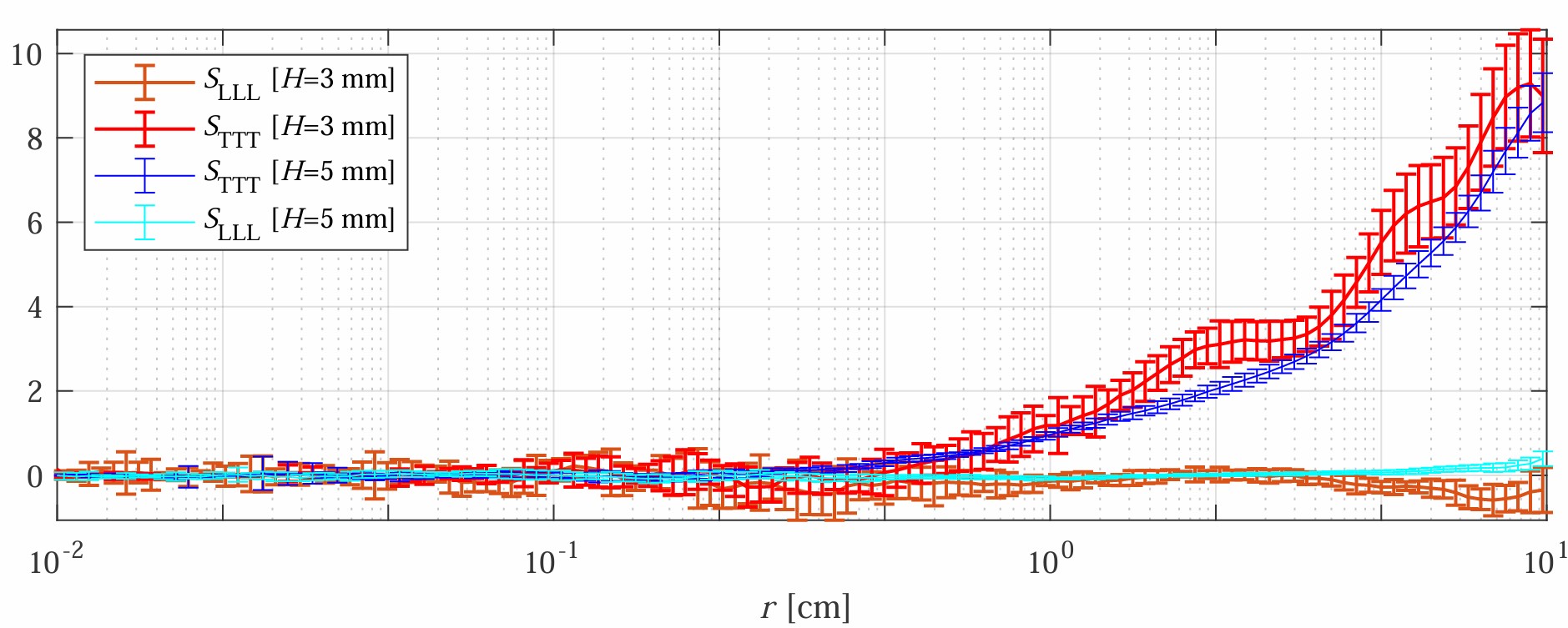}}
\caption{$S_{LLL}$ and $S_{TTT}$ as a function of $r$ for the two studied heights. The error bars are estimated using the 10 different movies recorded with the same experimental conditions.}
\label{fig:SLLLSTTT3mm} 
\end{figure}

Kolmogorov derived the `4/5-law' for the three-dimensional turbulent flows 
\begin{equation}
S_{LLL}= \langle (\delta u_L)^3 \rangle = -\frac{4}{5} \, \varepsilon \, r ,    
\end{equation}
where $\varepsilon$ is the average energy dissipation rate per unit mass. The minus sign indicates the energy flux from large to small scales, which is the reason it is called a direct cascade. 

In 2D flows~\citep{lindborg1999can,cerbus2017third}, two cascades could co-exist leading to  
\begin{gather}
S_{LLL} \sim r^3, \; \hbox{in the direct cascade range, and } \\
S_{LLL} \sim r, \hbox{in the inverse cascade range }
\end{gather}
In the inverse cascade range, the KK theory predicts that the same type of scaling properties in the longitudinal and transverse directions as a function of $r$.  

\citep{belmonte1999velocity} investigated grid turbulence in a soap film and observed that the third moment is slightly negative at small scales but turns positive over most of the measured range. This is the signature of an energy transfer dominated by the inverse cascade. In a separate study, 
\citep{gledzer2011structure,gledzer2013effect} employed a $40 \times 30$~cm$^2$ container filled with an electrolyte solution up to 7~mm height subject to the Lorentz force. Their measurements revealed a negative third-order longitudinal structure function, consistent with a direct energy cascade akin to 3D turbulence. 

Fig.~\ref{fig:SLLLSTTT3mm} presents the third-order moments in both longitudinal ($S_{LLL}$) and transverse ($S_{TTT}$) directions for the heights of 3 and 5~mm, plotted on a semi-logarithmic scale. The second-order structure function indicates strong correlations at all scales. The third-order moment in the longitudinal direction remains close to 0 up $r \approx L\approx 2$~cm, after which it is slightly negative, suggesting a direct energy cascade. The fact that $S_{LLL}$ is close to 0 for $r<L$ could indicate that in this range, energy transfer is dominated by the enstrophy cascade. In the transverse direction, we note that the amplitude of $S_{TTT}$ is much greater than $S_{LLL}$, and it takes positive values starting from $r\approx H$. One could deduce that the inverse enstrophy cascade dominates the transfer, with this process starting to be important from $r=H$ and not from the distance among the magnets. 

Our analysis reveals three distinct regimes:
\begin{itemize}
 \item For separations $r<H$, the third-order structure function tends toward zero in both directions, despite the presence of pronounced fluctuations at these scales. Given the simultaneous presence of energy and enstrophy cascades in two-dimensional turbulence, this behavior may suggest that the interscale energy transfer is mediated predominantly through the enstrophy cascade.
 \item In the range $H>r>L$, the sign of $S_{TTT}$ is positive, while $S_{LLL}$ remains close to 0. Thus, one may deduce that the energy transfer starts to be important in the transverse direction, while in the longitudinal direction, the enstrophy continues to dominate. 
 \item For $r>L$, the amplitude $S_{TTT}$ continues to increase with positive values according to $r^1$ that indicate an inverse energy cascade, while $S_{LLL}$ remains  close to zero.
\end{itemize}

\begin {figure}[h]
\centerline{\includegraphics[scale=0.4]{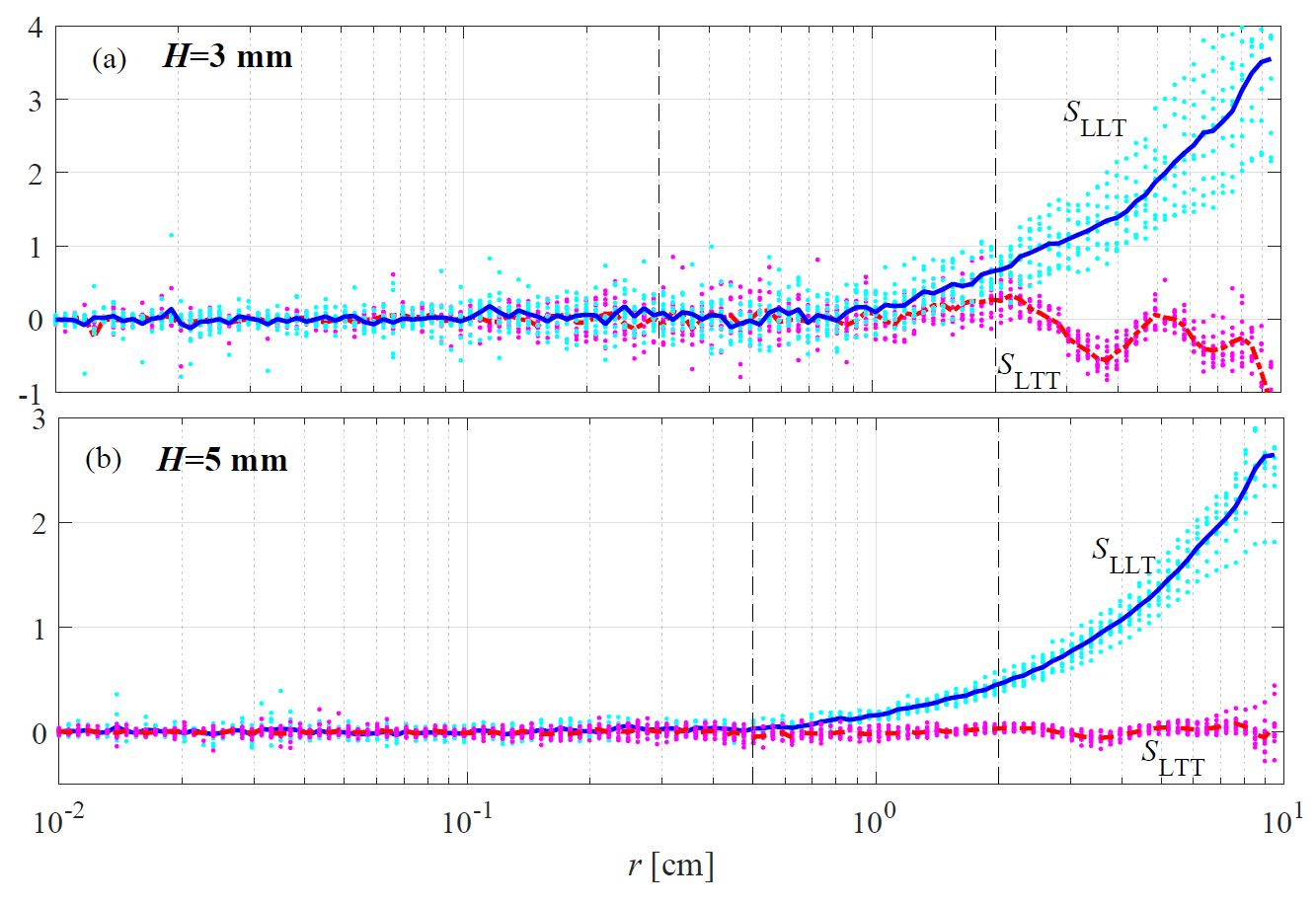}}
\caption{The off-diagonal third-order structure functions $S_{LLT}$ and $S_{LTT}$ as a function of $r$ for $H=3$ in (a) and 5~mm in (b). The dots are the results obtained for each movie, and the solid lines are the average over all the data. The cross-direction coupling is found to be above the experimental error at large scales.}
\label{fig:cross} 
\end{figure}

\subsection{Cross third-order structure functions}

Local anisotropy is detected at small scales, with different behaviors of the second-order structure function reported in the longitudinal and transverse directions. To investigate the correlation between the transverse and longitudinal directions, we use the cross third-order structure functions, 
\begin{equation}
S_{LTT} = \langle \delta u_T^2 \delta u_L \rangle \hbox{ and }
S_{LLT} = \langle \delta u_L^2 \delta u_T \rangle .    
\end{equation}
These functions are cross-third-order moments, which reflect the energy transfer among the scales in different directions. They enable us to assess the coupling between the two directions, {\em i.e.}, longitudinal and transverse.

In Fig.~\ref{fig:cross}, we plot $S_{LTT}$ and $S_{LLT}$ for the two heights as a function of $r$. The cross-correlation between the two directions is close to zero for the small scales and becomes important at large scales. For $S_{LLT}$, one can verify that it exhibits positive values for $r>H$ while remaining around 0 for $r<H$. The longitudinal and transverse directions are thus coupled mainly at large scales. The fact that for $r<H$, the cross-correlation is close to 0 is also obtained for $S_{LTT}$, which is found to be slightly negative with values above the experimental error for $r>H$ in agreement with the behavior of $S_{LLL}$.

We deduce that a net coupling among the two directions in the moving frame of reference is detected at $r>H$. Consequently, one could hypothesize the effects of bottom friction, which mainly affect the longitudinal direction, may be responsible for the deviation of the scaling in the transverse direction due to this nonlinear coupling at large scales between the two directions.  

\section{Conclusion} \label{sec:concl}

We report an experimental investigation of the dynamics obtained in a forced quasi-two-dimensional flow in a square container with a no-slip bottom boundary. The working fluid is a potassium hydroxide (KOH) electrolyte solution, with fluid layer heights of 3 mm or 5 mm. Complex motion is driven by an electric current imposed via electrodes on two sidewalls, coupled with a magnetic field generated by an array of permanent magnets (alternating polarities) placed beneath the container. Flow visualization is achieved using 50~$\mu$m fluorescent tracer particles seeded on the free surface. These particles absorb UV light (356~nm wavelength, provided by UV lamps) and emit in the visible spectrum. A high-resolution camera (50~$\mu$m/pixel, matching the particle diameter) captures the full domain, with careful attention to temporal analysis being restricted to periods of statistical stationarity. 

We show that for the currents used, which are 500 and 700~mA, respectively for the 3~mm and 5~mm heights, the velocity field becomes random and chaotic in space with the absence of the coherent vortices that are visible at low currents. However, at the Reynolds numbers investigated here, we cannot rule out the existence of coherent transitional structures. The randomness is also reflected in the PDF of the velocity, which is close to a Gaussian with skewness and flatness factors equal to 0 and 2.8, respectively.  

We compare the statistical properties obtained here with the KK theory. The deviation from the KK theory is observed through the PDFs of velocity increments, which not only exhibit non-Gaussian behavior but also differ between longitudinal and transverse motions relative to the moving frame of reference. Although the second-order structure function appears isotropic in the laboratory frame, this isotropy breaks down in the moving frame. By introducing a local anisotropy factor, we found that it peaks within the Q2D range, situated between the Kolmogorov scale and the flow height. Analysis of the third-order structure function highlights the prevalence of an inverse energy cascade, suggesting that small-scale local anisotropy is driven to larger scales. As a result, the observed scaling laws deviate from the predictions of KK theory in this direction. Furthermore, cross third-order structure functions reveal significant mixing between longitudinal and transverse motions, indicating strong correlations at large scales and an associated ``pollution'' effect spreading to the transverse direction.

One possible interpretation of our experimental results is as follows: The bottom no-slip boundary allows friction to act on the longitudinal direction of motion. This inherently breaks down the local isotropy of the flow. Larger scales are affected by the inverse enstrophy cascade. The transverse properties of turbulence are also affected because of their nonlinear coupling with the longitudinal direction. Consequently, the effects of the solid boundary would propagate in the two directions and make, at all scales, the statistical properties of turbulence in disagreement with the KK theory. In addition, because all of this study takes place at moderate Reynolds numbers, the possible existence of transitional coherent structures can also play a major role in the disagreement between the experimental data and the KK theory. 

For future work, it would be important to understand the theoretical reasons behind the dominant inverse cascade in these flows. Finally, let us mention that a comparison with direct numerical simulations yields another interesting perspective of this experimental work and will certainly lead to further detailed insight into the flow structure and the role of the solid boundary in modifying the flow's dynamics, especially when compared to theory.

\section*{Acknowledgments}
{SB and KS acknowledge the financial support from the French Federation for Magnetic Fusion Studies (FR-FCM) and the Eurofusion consortium, funded by the Euratom Research and Training Programme under Grant Agreement No. 633053. The views and opinions expressed herein do not necessarily reflect those of the European Commission. GA and JK acknowledge the financial support of the University Research Board (URB) award number 104391, and the Undergraduate Research Experience (URE) award number 11.513117.
}

\bibliography{Q2Dliterature}

\end{document}